\documentclass[usenatbib]{mnras}
\usepackage{lineno}
\usepackage{natbib}
\usepackage{psfig}
\usepackage{graphicx}
\usepackage{amssymb}
\usepackage{color}
\usepackage{setspace}

\newcommand{\ptrsa}{RSPTA}
\newcommand{\cursci}{Curr. Sci.}
\newcommand{\sa}{SvA}

\author[Perera et al.]{B.~B.~P.~Perera$^{1}$, B.~W.~Stappers$^{1}$, A.~G.~Lyne$^{1}$, C.~G.~Bassa$^{2}$, I.~Cognard$^{3,4}$,
\newauthor L.~Guillemot$^{3,4}$, M.~Kramer$^{5,1}$, G.~Theureau$^{3,4,6}$ , G. Desvignes$^5$
\\ $^1$ Jodrell Bank Centre for Astrophysics, School of Physics and Astronomy, The University of Manchester, Manchester M13 9PL, UK
\\ $^2$ ASTRON, the Netherlands Institute for Radio Astronomy, Postbus 2, 7990 AA, Dwingeloo, the Netherlands
\\ $^3$ Laboratoire de Physique et Chimie de l'Environnement et de l'Espace LPC2E CNRS-Universit\'{e} d'Orl\'{e}ans, F-45071 Orl\'{e}ans, France
\\ $^4$ Station de radioastronomie de Nan\c{c}ay, Observatoire de CNRS/INSU, F-18330 Nan\c{c}ay, France
\\ $^5$ Max-Planck-Institut f$\ddot u$r Radioastronomie, Auf dem H$\ddot u$gel 69, D-53121 Bonn, Germany
\\ $^6$ Laboratoire Univers et Th\'{e}ories LUTh, Observatoire de Paris, CNRS/INSU, Universit\'{e} Paris Diderot, 5 place Jules Janssen, 
\\F-92190 Meudon, France
}

\title[Revealing the centre of NGC 6624 through PSR B1820$-$30A]{Evidence for an intermediate-mass black hole in the globular cluster NGC 6624}

\begin{document}

\maketitle


\begin{abstract} 
PSR B1820$-$30A is located in the globular cluster NGC 6624 and is the closest known pulsar to the centre of any globular cluster. We present more than 25 years of high-precision timing observations of this millisecond pulsar and obtain four rotational frequency time derivative measurements. Modelling these higher-order derivatives as being due to orbital motion, we find solutions which indicate that the pulsar is in either a low-eccentricity ($0.33\lesssim e\lesssim0.4$) smaller orbit with a low mass companion (such as a main sequence star, white dwarf, neutron star, or stellar mass black hole) or a high-eccentricity ($e\gtrsim0.9$) larger orbit with a massive companion. The cluster mass properties and the observed properties of 4U 1820$-$30 and the other pulsars in the cluster argue against the low-eccentricity possibility. 
The high-eccentricity solution reveals that the pulsar is most likely orbiting around an intermediate-mass black hole (IMBH) of mass $> 7,500$~M$_\odot$ located at the cluster centre. 
A gravitational model for the globular cluster, which includes such a central black hole (BH), predicts an acceleration that is commensurate with that measured for the pulsar. 
It further predicts that the model-dependent minimum mass of the IMBH is $\sim60,000$~M$_\odot$. Accounting for the associated contribution to the observed period derivative indicates that the $\gamma$-ray efficiency of the pulsar should be between 0.08 and 0.2. Our results suggest that other globular clusters may also contain central black holes and they may be revealed by the study of new pulsars found sufficiently close to their centres. 

Note that we found an erratum in Section 5 and thus, the $\sim$60\,000~M$_\odot$ mass mentioned above has to be replaced by the correct model-dependent mass limit of $\sim$20\,000~M$_\odot$. See the erratum appended.

\end{abstract}



\section{Introduction}

Globular clusters contain large numbers of old stars gravitationally bound together in regions of a few tens of light years across. High stellar densities towards the centre of globular clusters provide a likely environment for the formation of massive BHs \citep{pbh+04,mh02,grh05,lkg+13}. The dynamics of the inner region are dominated by the presence of any central BH and the motion of stars around the BH can potentially be measured. Millisecond pulsars (MSPs) are one of the most stable rotators in the universe and their stability of the rotation is comparable to that of an atomic clock \citep[e.g.][]{pt96,hcm+12}. They are therefore very sensitive to any dynamical changes caused by the presence of a central BH. They are old neutron stars that are spun up to millisecond periods during a mass accretion phase through a so-called recycling process \citep{rs82,acrs82}.  The vast majority of the known pulsars in globular clusters are MSPs -- about 120 MSPs discovered in 28 globular clusters\footnote{http://www.naic.edu/$\sim$pfreire/GCpsr.html}.

\begin{figure}
\includegraphics[width=7cm]{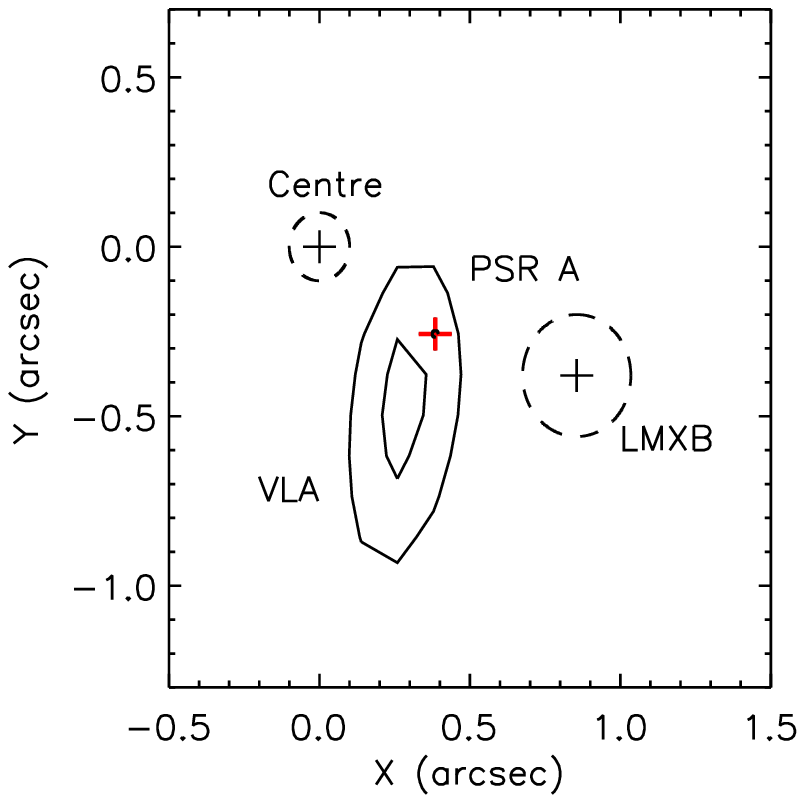}
\caption{
Radio sources in the central region of the globular cluster NGC 6624. The best available position measurement of PSR B1820$-$30A \citep{lfrj12} is marked with the {\it red} cross and that of LMXB 4U 1820$-$30 \citep{mfr+04} and the cluster centre \citep{gra+10} are marked with {\it black} crosses. The {\it dashed circles} represent the measured uncertainties of these source locations and that of the cluster centre. The contours are the 80\% and 95\% intensity levels of the VLA radio continuum region \citep{mfr+04}. Note that the other pulsars in the cluster are located outside of this field-of-view (i.e. the projected distances of pulsar B, C, and D are $\gtrsim4''$, while pulsar E and F do not have precise position measurements$^\dagger$). 
}
$^\dagger$\scriptsize See ATNF pulsar catalog (http://www.atnf.csiro.au/people/pulsar/psrcat/) 
\label{cluster}
\end{figure}

The Milky Way globular cluster NGC 6624 is located 7.9~kpc away from the Earth, and 1.2 kpc away from the Galactic centre based on optical observations \citep[][2010 edition]{har96}. \citet{kdi+03} estimated the distance to the cluster to be 7.6(4)~kpc based on Type I X-ray bursts from the source 4U 1820$-$30. This Galactic bulge globular cluster is thought to be core-collapsed due to the cusp signature seen in the density profile derived from optical observations \citep{sk95,ng06}. Six radio pulsar have been discovered so far in NGC 6624; PSRs B1820$-$30A and B \citep[PSRs J1823$-$3021A \& B][]{blma90,bbl+94}, and PSRs J1823$-$3021C, D, E, and F \citep{cha03,lfrj12}. PSRs  B1820$-$30B and J1823$-$3021C are young pulsars with a spin period of $\sim 0.4$~s, while the others are MSPs with a spin period of $<6$~ms. The position measurements of these pulsars are given in Table~\ref{position}. PSR B1820$-$30A is the closest known pulsar to the centre of any globular cluster, where its projected separation from the cluster centre in the sky-plane is approximately $0.5''$ (see Figure~\ref{cluster}). It has a rotational frequency ($f$) of 183.82~Hz and frequency time derivative ($\dot f$) of $-1.14\times10^{-13}$~Hz$/$s \citep{bbl+94,lfrj12}, where the period derivative is $\dot{P} = -\dot{f}/f^2 = 3.38\times10^{-18}$~s$/$s. The values of $\dot P$ and $\dot f$ are usually attributed to the loss of rotational energy from the spinning neutron star. However, this is the highest measured $\dot P$ for any MSP and is several orders of magnitude greater than the typical value \citep[see][]{mhth05}\footnote{http://www.atnf.csiro.au/people/pulsar/psrcat/ }. It has previously been proposed that this anomalous $\dot P$ is not intrinsic to the pulsar and is solely induced by the dynamics due to its special location in the cluster \citep{bbl+94}. However, the derived high luminosity from the observed pulsed $\gamma$-ray flux density of the pulsar using the \textit{Fermi} LAT suggests that a notable fraction of the observed value of $\dot P$ could be intrinsic \citep{faa+11,aaa+13}. In general, the intrinsic $\dot P$ of pulsars can be estimated from the observed gamma-ray luminosity, although the unknown $\gamma$-ray beaming fraction, $\gamma$-ray efficiency, and the large uncertainties of distance estimates make them poorly constrained for individual objects.

\begin{table}
\begin{center}
\caption{
The available position measurements of the cluster centre, radio pulsars, and LMXB 4U 1820$-$30. Note that the accurate positions of PSRs J1823$-$3021E and F are not available \citep[see][]{lfrj12}.  
}
\label{position}
\begin{tabular}{lccc}
\hline
\multicolumn{1}{l}{Source} &
\multicolumn{1}{c}{RA} &
\multicolumn{1}{c}{DEC} &
\multicolumn{1}{c}{Ref} \\
\hline
Cluster centre & 18$:$23$:$40.510(7) & $-30$$:$21$:$39.7(1) & 1 \\
 & 18$:$23$:$40.55 & $-30$$:$21$:$39.6 & 2 \\
B1820$-$30A & 18$:$23$:$40.4871(4) & $-30$$:$21$:$40.13(4) & 3 \\
 & 18$:$23$:$40.48481(9) & $-30$$:$21$:$39.947(9) & 4 \\
 B1820$-$30B & 18$:$23$:$41.546(2) & $-30$$:$21$:$40.9(5) & 3 \\
 J1823$-$3021C & 18$:$23$:$41.152(4) & $-30$$:$21$:$38.4(8) & 3 \\
 J1823$-$3021D & 18$:$23$:$40.531(7) & $-30$$:$21$:$43.7(4) & 3 \\
 4U 1820$-$30 & 18$:$23$:$40.45(1) & $-30$$:$21$:$40.1(2) & 5 \\
\hline
\end{tabular}
\begin{tabular}{l}
\multicolumn{1}{l}{} \\
Ref: (1) \citet{gra+10}; (2) \citet{kps+13}; \\
(3) \citet{lfrj12}; (4) This work; (5) \citet{mfr+04}
\end{tabular}
\end{center}
\end{table}

The orbital parameters and masses of a pulsar in an orbit around a companion are traditionally determined by fitting Keplerian and post-Keplerian models to timing data covering at least a complete orbit \citep{lk05}. However, this method is applicable to short orbital period systems and cannot be used for wide-orbit systems where the orbital periods are longer than the data set. \citet{jr97} proposed a method for measuring the orbits and companion masses of such binary MSPs that uses the higher-order spin frequency derivatives obtained from timing measurements. The main assumption in this model is that all the frequency derivatives are dynamically induced and not intrinsic to the pulsar. We use our precise timing measurements of PSR B1820$-$30A in this model to constrain the pulsar orbit and its companion mass. 
In our analysis, we also consider a significant pulsar intrinsic spin-down contribution in the observed value as mentioned in \citet{faa+11} and re-constrain the pulsar orbit and its companion mass.

NGC 6624 also contains the extremely-short-orbital-period Low Mass X-ray Binary (LMXB) 4U 1820$-$30 which has an anomalously large orbital period derivative that also appears to be influenced by the cluster dynamics \citep{spw87,rmj+87}. Using more recent data, \citet{pbk+14} also proposed that the observed negative orbital period derivative of the LMXB is due to an acceleration in the cluster potential and suggested that the centre of the cluster most likely contained a large amount of dark remnants, or an  IMBH, or that there was a dark remnant located close to the LMXB itself.

Depending on the orbit and the mass of the companion given the close location of the pulsar to the centre, we can reveal the properties of the central source, an IMBH in this case, of the cluster NGC 6624. The possibility of MSPs orbiting around IMBHs in globular clusters has previously been discussed \citep[see][]{cmp03,dcm+07}. However, such a system has not been evident prior to this work. The best observational evidence hitherto for IMBHs is based on line broadening seen in the centres of dwarf galaxies \citep{rgg13} and the implied high mass accretion rates of the ultra-luminous X-ray sources \citep{ppd+06,fwb+09}. The current best candidates for globular clusters hosting central IMBHs are NGC 5139 \citep{ngk+10} and G1 in M 31\citep{grh05}, and NGC 6388 \citep{lgb+15} in the Milky Way based on  the velocity dispersion of stars near the cluster centres. However, in the latter case there is some disagreement on the mass ($\sim$$30,000$~M$_\odot$) of the IMBH \citep{lmo+13,lgb+15}. 
The IMBHs in dense star clusters are also expected to be detected by their accretion signatures in X-ray and radio continuum \citep{mac04}. Several studies searched for such emission, but could not find potential signatures. The centre of G1 was detected in both radio and X-ray \citep{ugh07,pr06}, but later observations found no such radio signature \citep{mws+12}.

The low fluxes in radio and X-ray measurements of some globular cluster centres lead to very low mass upper limits on possible central BHs \citep{scm+12a,ckc+10,lk11}. However, the various assumptions used may have affected the results \citep[see][]{sjg+13,lkg+13}. 
The quiescent stellar-mass BH candidates have mainly been found in globular clusters as low-luminosity accreting BHs \citep[see][and references therein]{msh+15}. Recently, such a BH was identified outside of a globular cluster for the first time \citep{tba+16}. The small gas content in globular clusters \citep[see][and references therein]{vsem06,fkl+01} perhaps makes the accretion onto the central IMBH weaker than expected, resulting in low or absent measured fluxes in radio continuum and X-ray. Although, we note that the dispersion measures of MSPs in the Globular cluster 47 Tucanae suggest that it contains some gas \citep{fkl+01}, implying that the accretion cannot be completely ignored in that particular cluster and perhaps others.
\citet{ogff97} interpreted the presence of the intracluster medium of this particular cluster as perhaps being due to a possible bow shock formed from the interaction of the cluster with the Galactic halo.  
In general, \citet{mz15} proposed that ultraviolet radiation from white dwarfs in globular clusters can efficiently ionise and eject the intracluster medium.

The plan of the paper is as follows. In Section~\ref{obs}, we present our observations, data processing, and the pulsar timing results. We study the possible acceleration terms that apply to the pulsar in Section~\ref{cluster_acc} and then establish the fact that the observed spin-down of the pulsar is dominated by dynamics, which is the main assumption of the orbit model given in \citet{jr97}. In Section~\ref{model}, we summarise this orbit model and then describe the method used to determine the orbital parameters and the masses.
We find the timing measurements are only consistent with the pulsar being in either a high-eccentricity larger orbit with a massive companion, an IMBH in this case, or in a low-eccentricity smaller orbit with a less massive companion.      
We present the high-eccentricity pulsar orbit solution in Section~\ref{around_centre} and argue that it is the most probable scenario. 
We also then explore the possibility of the pulsar existing in a low-eccentricity orbit in Section~\ref{small} and rule out some of these solutions based on the properties of the cluster. In Section~\ref{dyn}, we use a gravitational model of globular clusters including a central IMBH to derive the acceleration of sources around the cluster centre and then compare the results with their measured values to investigate the BH in more details. Finally in Section~\ref{dis}, we discuss our results.

\section{Observations and timing PSR B1820$-$30A}
\label{obs}

The timing observations of PSR B1820$-$30A were obtained with the Lovell Telescope (LT) at the Jodrell Bank Observatory in the UK since its discovery in 1990 March. These observations were made roughly every 18 days over more than 25 years, providing the largest ever data set available for this pulsar,  resulting in a total of 516 epochs. The observations were mainly carried out at L-band frequency (the centre frequency varies between 1400 and 1520 MHz) combined with some early 600~MHz observations. Two pulsar backends were used to record the data through out the entire observation span: the `analog filter bank' \citep[AFB;][]{sl96} used from 1990 March to 2010 April and the `digital filter bank' (DFB) used since 2009 October. In addition to the LT observations, the pulsar was observed using the Nan\c{c}ay Radio Telescope (NRT) in France since 2006 February, resulting in a total of 61 epochs. These NRT observations were made at L-band frequency ($\sim$1400~MHz) and the data were recorded using two backends: `Berkeley-Orl\'{e}ans-Nan\c{c}ay' \citep[BON;][]{ct06} backend with a frequency bandwidth of 64 MHz before July 2008 and 128 MHz after, and using the `NUPPI' backend \citep{ctg+13} with a bandwidth of 512 MHz after it became the principal instrument for pulsar observations in August 2011.

The AFB data were processed using the pulsar timing program `PSRPROF'\footnote{http://www.jb.man.ac.uk/pulsar/observing/progs/psrprof.html}, while the DFB, and NRT BON and NUPPI data were processed using the pulsar data processing package `PSRCHIVE'\footnote{http://psrchive.sourceforge.net}. For each observation, we folded the data in time (i.e. across the observation length) and frequency (i.e. across the bandwidth of the backend) using the previously published pulsar timing solution to obtain a high signal-to-noise pulse profile \citep{hlk+04}. We then use this pulse profile with the backend-dependent noise-free template of the pulsar to obtain the pulse time-of-arrival (TOA) for the given observation epoch \citep{hem06}. Combining both LT and NRT observations, we analysed 577 TOAs in total in the timing analysis.

\begin{figure}
\includegraphics[width=8cm]{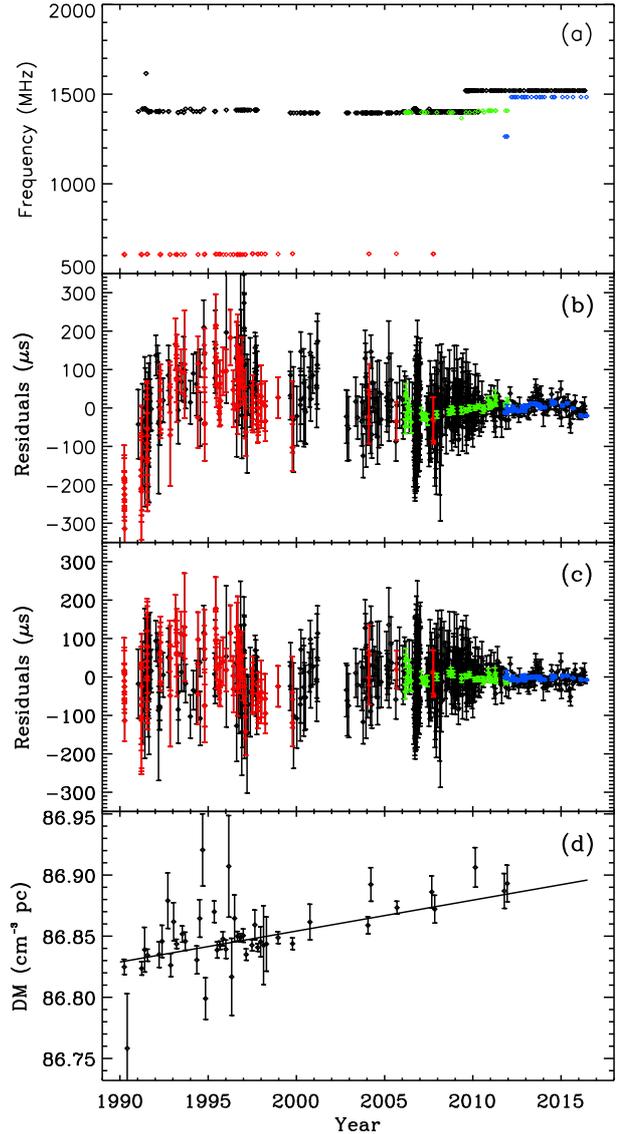}
\caption{
The observations at different observing frequencies as a function of time are shown in panel (a) -- LT 1400 MHz ({\it black}), LT 600 MHz ({\it red}), NRT BON ({\it green}), and NRT NUPPI ({\it blue}). Panel (b) shows the residuals of the pulsar after fitting only for the first two rotational frequency derivatives (i.e. $\dot f$ and $\ddot f$) with other parameters given in Table~\ref{solution}. Panel (c) shows the same as panel (b), but after fitting for all four rotational frequency derivatives. 
Panel (d) shows the time-dependent DM values constrained by using the method given in \citet{kcs+13}. The straight line represents the linear least-squares fit to the data points.
}
\label{res_plot}
\end{figure}

We fit a standard pulsar timing model including the astrometric parameters, the dispersion measure DM (i.e. the frequency-dependent time delay in the pulsar emission due to electron density in the interstellar medium along the line-of-sight), and the rotational frequency parameters to obtain the residuals between the observed and model predicted TOAs using the pulsar timing software TEMPO2 \citep{hem06}. When combining TOAs from different telescopes and pulsar backends, a time offset or `JUMP' in the model was used to account for any systematic delays between the data sets. We began the fit by only including the rotational frequency $f$ and its first time derivative $\dot f$ in the model and found a large remaining structure in the residuals. To minimise these residuals, we found that it is necessary to include higher order frequency derivatives, up to the fourth order (i.e. $f^{(4)}$), where the resultant accuracy of $f^{(4)}$ was 2.2$\sigma$. The timing residuals are shown in Figure~\ref{res_plot} and the timing solution is given in Table~\ref{solution}. We also fit for $f^{(5)}$ and found that its uncertainty is large, so that it provides only an upper limit. In the future with more observations, we will be able to constrain its value better. The best-fit timing model calculates the TOA-uncertainty-weighted root-mean-square of the residuals to be about 12~$\mu$s, which is about a factor of 12 improvement compared to the previous analysis given in \citet{hlk+04} that was based on 12.7 years of LT observations including only 275 TOAs.

Using the early 600~MHz AFB observations with the LT combining with other L-band observations (with different centre frequencies between $\sim1400 -1520$~MHz), we fit for the dispersion measure DM of the pulsar. We further fit for its first time derivative $\dot{\rm DM}$ to model any existing time-dependent variation in DM (see Table~\ref{solution}). Our measurements are consistent with the previously published values \citep[see][]{hlk+04}. The DM and $\dot{\rm DM}$ is less constraining within the L-band-only observation period due to a narrower frequency coverage, in particular after year 2012 (see the {\it top} panel in Figure~\ref{res_plot}). 
However, we assume that there is no significant variation in DM after year 2012 and thus, use the earlier $\dot{\rm DM}$ given in Table~\ref{solution} across the entire observation period. For comparison, we use the method introduced in \cite{kcs+13} to investigate the DM variation as a function of time. We fit this particular DM model with our timing model parameters, excluding $\dot{\rm DM}$. The {\it bottom} panel in Figure~\ref{res_plot} shows the model-estimated DM values as a function of time. A simple linear fit to these changes calculates a $\dot{\rm DM}$ of $2.5(3)\times10^{-3}$~cm$^{-3}$~pc~yr$^{-1}$, which is equal to the value given in our timing model (see Table~\ref{solution}) within the errors. 
We further note that the measured higher-order rotational frequency derivatives when this particular DM model is included in the timing model are consistent with our measurements given in Table~\ref{solution}.

According to \citet{kps+13}, the proper motion of NGC 6624 is small; (PMRA, PMDEC$) = (-1.29, -9.77)$~mas/yr with an uncertainty of 1.06~mas/yr. The proper motion of the pulsar measured from timing (see Table~\ref{solution}) is also small and similar to that of the cluster. Therefore, the relative motions of the cluster in the sky with respect to the Earth and the pulsar within the cluster are small, indicating that the change in our line-of-sight is small, and thus, not sensitive to larger scale changes in the interstellar medium. Therefore, we include the earlier $\dot{\rm DM}$ in the timing solution and assume that there has been no significant change since the low-frequency observation stopped.

\begin{table*}
\begin{center}
\caption{
Timing model parameters of PSR B1820$-$30A. The number in parentheses is the 1-$\sigma$ uncertainty in the last quoted digit.
}
\label{solution}
\begin{tabular}{lc}
\hline
\multicolumn{1}{l}{Timing parameters} &
\multicolumn{1}{c}{} \\
\hline
Right ascension RA (J2000) & 18:23:40.48481(9) \\
Declination DEC (J2000) & $-$30:21:39.947(9)\\
Proper motion in RA, PMRA (mas$/$yr) & $-$0.1(1) \\
Proper motion in DEC, PMDEC (mas$/$yr) & $-7(1)$ \\
Spin frequency $f$ ($\rm s^{-1}$) & 183.823411511659(7) \\
Spin frequency $\rm 1^{st}$ derivative $\dot f$ ($\rm s^{-2}$) & $-1.1425447(7)\times10^{-13}$ \\
Spin frequency $\rm 2^{nd}$ derivative $\ddot f$ ($\rm s^{-3}$) & $5.441(6)\times10^{-25}$ \\
Spin frequency $\rm 3^{rd}$ derivative $f^{(3)}$ ($\rm s^{-4}$) & $5.1(6)\times10^{-35}$ \\
Spin frequency $\rm 4^{th}$ derivative $f^{(4)}$ ($\rm s^{-5}$) & $1.5(7)\times10^{-43}$ \\
Spin frequency $\rm 5^{th}$ derivative $f^{(5)}$ ($\rm s^{-6}$)$^\dagger$  & $1(2)\times10^{-52}$ \\
Reference epoch of position (MJD) & 53000.0 \\
Reference epoch of period (MJD) & 52451.0 \\
Reference epoch of Dispersion measure (MJD) & 52451.0 \\
Dispersion measure DM (cm$^{-3}$~pc) & 86.852(1) \\
Dispersion measure $\rm 1^{st}$ derivative ${\rm \dot{DM}}$ (cm$^{-3}$~pc~yr$^{-1}$) & $2.9(2)\times10^{-3}$ \\
\\
Data span (MJD) & 47977 -- 57536 \\
Number of TOAs & 577 \\
Weighted rms timing residual ($\mu$s) & 12.32 \\
Reduced $\chi^2$ value & 1.09 \\
\hline 
Orbital dynamics model constrained parameters$^{\dagger\dagger}$ & $ $ \\
\hline 
Eccentricity e & 0.993(4)\\
Longitude of periastron $\omega$ ($^\circ$) & $353(2)$ \\
Longitude of the reference epoch $\lambda$ ($^\circ$) & $182.9(7)$ \\
Orbital period $P_{\rm orb}$ (kyr) & 1.6(4) \\
\hline
\end{tabular}
\begin{tabular}{l}
\multicolumn{1}{l}{} \\
$^\dagger$All parameters, except $f^{(5)}$, were fitted simultaneously. The limit of $f^{(5)}$ was obtained \\by fitting for all parameters while keeping $f^{(3)}$ and $f^{(4)}$ at their best values. \\
$^{\dagger\dagger}$Assuming the pulsar intrinsic spin-down is small compared to the observed value \\induced by dynamics (see Section~\ref{result_a}). \\
\end{tabular}
\end{center}
\end{table*}

Previous timing measurements spanning a smaller data set only revealed up to a maximum of two rotational frequency derivatives of the pulsar \citep{hlk+04}. The timing solution reported in \citet{lfrj12} was able to obtain frequency measurements only up to $\dot f$ using the Green Bank Telescope, because their observations spanned less than a year. 
We note that only one MSP (PSR J1024$-$0719) has been shown to require higher than a second order frequency derivative to model its arrival times and \citet{bjs+16} recently showed that these higher-order derivatives are dynamically induced by its binary motion in a wide orbit and not intrinsic to the source.
Note that the measured second order rotational frequency derivative of PSR B1820$-$30A is several orders of magnitude greater than that of other MSPs in general \citep{mhth05}. Since MSPs are stable rotators, we do not expect to see such large and higher order spin-down as we observe in PSR B1820$-$30A. Therefore, these higher order derivatives are not likely to be intrinsic to the pulsar, rather dynamically induced due to binary motion.

\section{Acceleration of the pulsar}
\label{cluster_acc}

The observed spin period derivative ($\dot P$) of a globular cluster pulsar is somewhat different to its intrinsic value due to various acceleration terms. In general, the measured value is affected by accelerations due to differential Galactic potential ($a_G$), toward the Galactic plane ($a_z$), cluster potential ($a_c$), and the proper motion of the pulsar across the sky \citep[so-called Shklovskii effect --][]{shk70}. As given in previous studies \citep[see,][]{phi92,nt95,bjs+16}, this can be expressed in the form  

\begin{equation} 
\label{accel}
\frac{\dot{P}}{P} \approx \frac{\dot{P}_{\rm int}}{P} + \frac{a_c}{c} + \frac{a_G}{c} + \frac{a_z}{c} + \frac{\mu^2D}{c}
\end{equation}

\noindent
where $P$ and $\dot{P}_{\rm int}$ are the observed period and the first time derivative of the intrinsic period, $\mu$ is the proper motion, and $D$ is the distance to the pulsar. For PSR B1820$-$30A, the calculated differential Galactic acceleration $a_G/c \approx -5.7\times10^{-11}$~yr$^{-1}$ given in \citet{pbk+14}, the acceleration toward the Galactic plane $a_z/c \approx 4.56\times10^{-12}$~yr$^{-1}$ \citep[see Equation 4 in][]{nt95}, and the Shklovskii effect $\mu^2D/c \approx 5.89\times10^{-11}$~yr$^{-1}$ for the pulsar distance of $D=7.6$~kpc  \citep{kdi+03}. These terms are all negligible compared to the observed $\dot{P}/P$ and given that the typical spin period of $\sim (1.4 - 10)$~ms and intrinsic spin period derivative of $\sim 10^{-20}$~s/s of the MSP population the intrinsic contribution ($\dot{P_{\rm int}}/P$) could be expected to be comparably small ($\sim$$10^{-10}$~yr$^{-1}$). In this scenario we can equate the measured value to be equivalent to an acceleration in the potential of the cluster, i.e.  $a/c = \dot{P}/P = 1.960117(1)\times10^{-8}$~yr$^{-1}$.  This high acceleration already suggests that it could be induced by the orbital motion of the pulsar.  Thus, we apply the model proposed in \citet{jr97} to the timing data of PSR B1820$-$30A to determine all possible orbital system parameters and the companion masses consistent with the higher order frequency derivatives.  We do also note that the $\gamma$-ray observations of the pulsar suggest a high $\gamma$-ray luminosity that would require an intrinsic period derivative comparable to that of the observed value \citep{faa+11}. Thus, we also consider the influence of  a significant intrinsic spin period derivative of the pulsar in the orbit model in Section~\ref{result_b} and calculate re-constrained orbital parameters.

\section{Determining the orbital parameters and masses}
\label{model}

The method given in \citet{jr97} enables the determination of the orbital parameters of a binary pulsar and its companion mass up to the unknown factor $\sin(i)$, where $i$ is the orbital inclination angle. We use the `four derivative' solution given in their study with our measured frequency derivatives given in Table~\ref{solution} to determine many parameters of the binary orbit. The derived analytical expressions for the model are expressed as a function of orbital parameters eccentricity $e$, the longitude of periastron $\omega$ (measured from the ascending node), the longitude of the reference epoch $\lambda$ (measured from the periastron), and its first time derivative $\dot \lambda$. These equations are given in Appendix~\ref{app}. We derived the expression for $f^{(5)}$ (Equation~\ref{f5}) in this work for the comparison between the model predicted value and the measured upper limit given in Table~\ref{solution}.

\begin{figure}
\begin{center}
\includegraphics[width=3.3in]{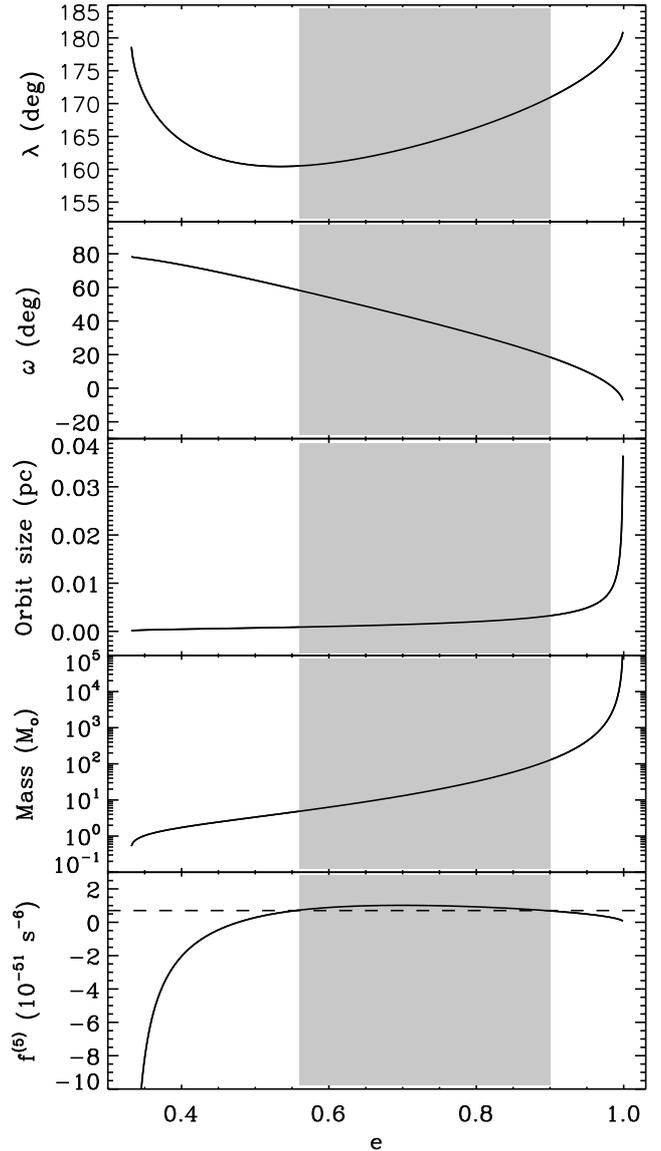}
\end{center}
\caption{
All possible pulsar binary systems for an edge-on orbit as a function of eccentricity. Note that there are valid solutions only when $e\gtrsim0.33$. The parameters $\lambda$ and $\omega$ are independent of the orbital inclination, while the derived pulsar orbit size (or the semi-major axis), and the companion mass are orbital-inclination-dependent parameters. For low eccentricities, the pulsar orbital size becomes small and the companion mass reduces and moves towards the sub-stellar regime. The {\it bottom} panel shows the orbital model-predicted ({\it solid} line) and the timing measured upper limit ({\it dashed} line) of $f^{(5)}$ for the pulsar. This indicates that when $0.56\lesssim e\lesssim0.9$ (i.e. the {\it shaded} region), the predicted $f^{(5)}$ is greater than its measured 3-$\sigma$ upper limit (i.e. $7\times 10^{-52}$~s$^{-6}$ -- see Table~\ref{solution}) and thus ruling out such binary systems. 
\label{other}}
\end{figure}

We first investigate all possible orbital systems of the pulsar based on its measured higher-order rotational frequency derivatives through timing. 
The system has three equations (Equation~\ref{f2}--\ref{f4}) with four unknowns ($e$, $\omega$, $\lambda$, and $\dot \lambda$). As mentioned in \citet{jr97}, assuming a value for one parameter, we can solve for the remaining parameters using the Newton-Raphson method. 
In order to reduce the number of variables, we slightly modified the standard equations of the model and obtained Equation~\ref{f3_new} and \ref{f4_new}. These two expressions contain only three unknowns $\lambda$, $\omega$, and $e$. 
We first solve for orbital-inclination-independent parameters $\lambda$ and $\omega$ as a function of $e$ using these expressions. Our results are shown in Figure~\ref{other}. Since these are non-linear systems, more than one solution may exist. Thus, we searched for solutions using the initial guesses for the parameters across their entire space (i.e. [0,2$\pi$) for $\lambda$ and $\omega$, across [0,1) for $e$) and then use the best possible solution. We find solutions only when $e$$\gtrsim$$0.33$. 
Then we use these orbital parameters to determine the pulsar orbit size (Equation~\ref{asin}), or the semi-major axis of the orbit, and the mass of the companion (Equation~\ref{msin}) as a function of $e$ up to the given orbital inclination. Figure~\ref{other} shows the lower limits of these two parameters assuming an edge-on orbit (i.e. $i=90^\circ$)

To confirm our results we use the independent derivation of the frequency derivatives due to orbital motion as given by \citet{bjs+16}. By choosing random values for $e$, $\lambda$, and $\omega$, combined with the constraints on $\dot{f}$ and $f^{(2)}$, we performed a Monte Carlo simulation to select those orbital parameters that reproduce $f^{(3)}$ and $f^{(4)}$ within their 1-$\sigma$ measured uncertainties as given in Table~\ref{solution}. We found that the orbital parameters constrained from this method are exactly consistent with the results those constrained using the Newton-Raphson method (as shown in Figure~\ref{other}).

The orbital solutions consistent with the first four measured frequency derivatives allow eccentricities in excess of $e=0.33$. All solutions place the system near apastron of the orbit ($\lambda$ between 160 and 180\degr), with the low eccentricity solutions having low stellar mass companions in smaller orbits of which the apastron is placed towards the observer, i.e. $\omega\approx90^\circ$. For solutions with higher eccentricities, the companion mass and hence the orbit size  increase. To keep the same observed frequency derivatives, the apastron moves from being positioned out of the plane of the sky ($\omega=90\degr$) to being positioned in the plane of the sky ($\omega=0\degr$).
In order to further constrain the orbital parameters we derived an expression for $f^{(5)}$ (see Equation~\ref{f5}). Using this expression, we could estimate the model-predicted $f^{(5)}$ for the range of orbital solutions (see the bottom panel in Figure~\ref{other}) and compare this with the  measured upper limit from the pulsar timing. 
The 3-$\sigma$ upper limit of $f^{(5)}$ (the dashed line in Figure~\ref{other}) rules out all orbits where the eccentricity lies in the range between $0.56-0.9$.
The remaining valid high-eccentricity binary systems ($e\gtrsim0.9$) correspond to a large pulsar orbit (with the unknown orbital inclination) which could pass around a massive companion located at the cluster centre. For an orbit of eccentricity $\approx$0.9, we find that the inclination must be $\lesssim48^\circ$ in order that the orbit pass around the centre.
We investigate the parameters of these high-eccentricity orbits in Section~\ref{around_centre} in detail.  The constraint on $f^{(5)}$ also allows a range of low-eccentricity ($0.33\lesssim e\lesssim0.56$) pulsar binary systems with smaller orbits and low-mass companions $\approx (0.5-4.9)$~M$_\odot$, which could be main-sequence or slightly evolved cluster stars, white dwarfs, neutron stars or stellar-mass BHs. We discuss these in Section~\ref{small} and show that some of them can be ruled out by considering the properties of the cluster, the other pulsars, and the LMXB.

\subsection{Pulsar orbiting around the cluster centre}
\label{around_centre}

Here we examine the parameters of the pulsar if it is in a high-eccentricity orbit where $e$$\gtrsim$$0.9$. 
As mentioned before and shown in Figure~\ref{other}, these orbits are large enough to pass around the cluster centre where there is a massive central object. Therefore, we can use the observed projected separation of the pulsar from the cluster centre as an additional measurement to further constrain its orbital parameters. 
The cluster centre is measured to be R.A. $ = 18$$:$$23$$:$$40.51$ and DEC. $ = -30$$:$$21$$:$$39.7$ with an uncertainty of $0.1'' $ \citep{gra+10}. The absolute accuracy of the reference frame of this measurement and the uncertainty due to frame ties might give a total uncertainty of $<$$0.1''$ \citep[see][and private communication with Goldsbury et al.]{scs+06,asb+08}. We considered an additional error of $0.1''$ in the centre measurement given in \citet{gra+10}, and found that it does not change the orbital parameters and thus, the results presented in this work. Based on the measured cluster centre position given in \citet{gra+10} and \citet{kps+13} and the pulsar position given in \citet{lfrj12} and our timing solution (see Table~\ref{position}), the observed projected separation is calculated to be between $0.016$~pc $ \leqslant R_\perp \leqslant 0.042$~pc.

Using first four frequency derivatives in Equations~\ref{f2}--\ref{f4}, we determine the orbital parameters $e$, $\omega$, and $\lambda$ as a function of $\dot \lambda$.  
These values can then be used in Equations~\ref{r_perp}--\ref{zeqn} to calculate the model-derived projected separation of the pulsar from the cluster centre for an assumed edge-on orbit. We then compare this model-derived projected separation with the observed projected separation measurement given above (i.e. $0.016 - 0.042$~pc) to place a limit on $\dot \lambda$, and then keep that value fixed in the analysis. 
We note that the model-derived projected distance varies with the inclination angle, resulting in a change in the possible $\dot \lambda$ range. 
In general, the chance of observing a binary system at an angle $\lesssim26^\circ$ is about only 10\% \citep[see][]{lk05}.
Therefore, we consider all possible orbital inclinations $\gtrsim26^\circ$ and find out that the limit of $\dot \lambda$ varies slightly, and thus, our orbital results vary only by $<3$\%.
Therefore, we use the $\dot \lambda$ limit obtained from the edge-on orbit for any given orbital inclination in the model.

\subsubsection{ Case I: Orbital results assuming a negligible intrinsic spin-down of the pulsar}
\label{result_a}

\begin{figure}
\begin{center}
\includegraphics[width=3.in]{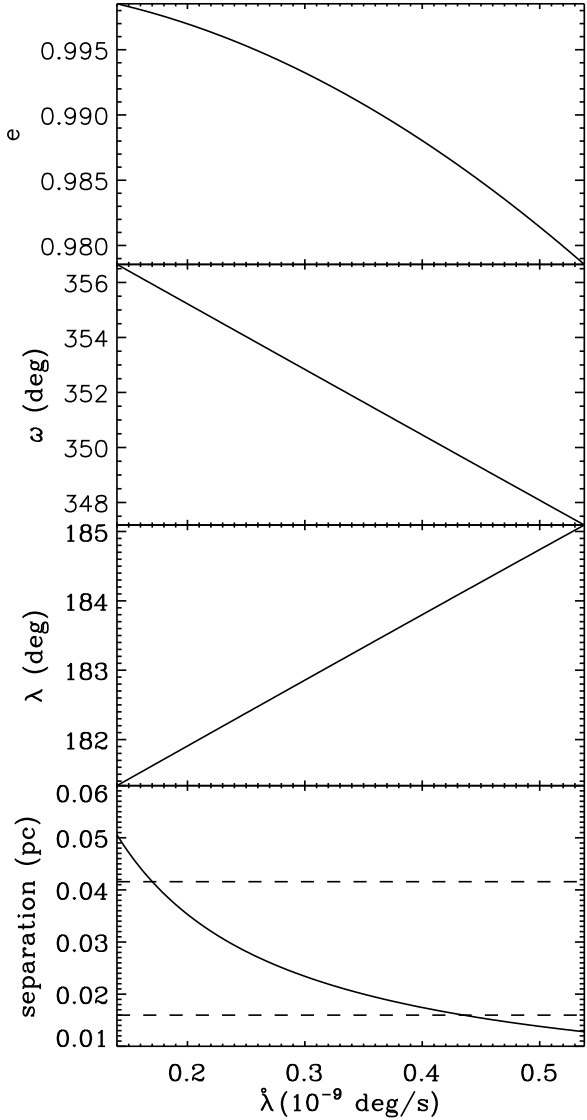}
\end{center}
\caption{
Solutions for orbital-inclination-independent parameters $e$, $\omega$, and $\lambda$ as a function of $\dot \lambda$. The {\it bottom} panel shows the projected separation of the pulsar from the cluster centre on the sky plane calculated based on $\omega$, $e$, and $\lambda$ as a function of $\dot \lambda$ for an edge-on orbit ($i=90^\circ$). The possible measured projected separation range ($0.016$~pc $ \leqslant R_\perp \leqslant 0.042$~pc) is marked by {\it dashed} lines. This implies the range of possible $\dot \lambda$ values to be $(1.7-4.3)\times10^{-10}$~deg/s.
\label{sol_lambda}}
\end{figure}

We first consider the case where the pulsar intrinsic spin-down is negligible compared to its observed value which is therefore assumed to be predominantly induced by dynamical effects (see Section~\ref{cluster_acc}). 
Our results for $e$, $\omega$, and $\lambda$ are shown in Figure~\ref{sol_lambda} where in the bottom panel we show the model-derived projected separation and the observed projected separation (dashed lines). These limits can then be used to impose a range of sensible values of $\dot \lambda$ to be $(1.7-4.3)\times10^{-10}$~deg/s. For the rest of this analysis we therefore use the mean value of the possible range of $\dot \lambda$ as the best value ($\dot \lambda_{\rm mean} = 3\times10^{-10}$~deg/s) and keep it fixed in the model. 
We further note that our orbital results vary only by less than 1\% within this possible $\dot \lambda$ range, and thus keeping $\dot\lambda (\equiv\dot \lambda_{\rm mean})$ fixed in the model is a valid assumption.

\begin{figure}
\begin{center}
\includegraphics[width=3.1in]{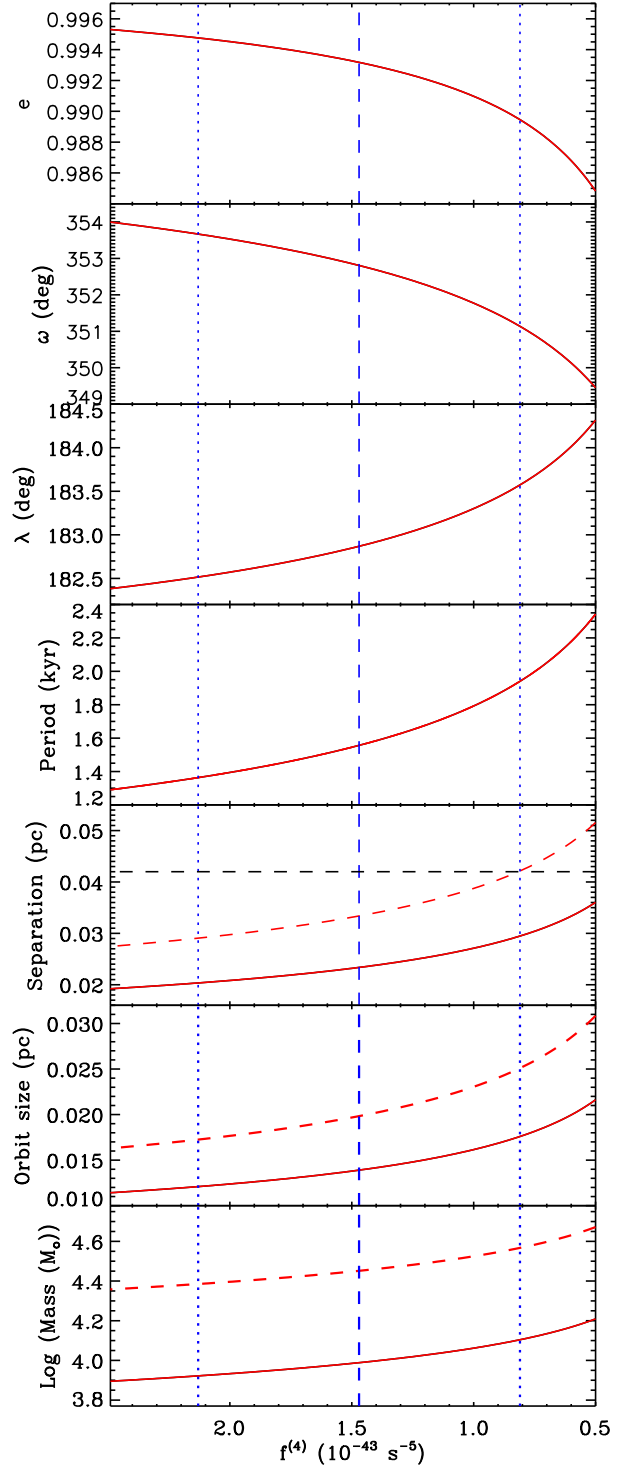}
\end{center}
\caption{
Parameters of the pulsar orbit around the cluster centre and the companion as a function of measured $f^{(4)}$. The vertical {\it blue dashed} and {\it blue dotted} lines represent the timing measurement of $f^{(4)}$ with its uncertainty. The minimum ({\it red solid}) and the maximum ({\it red dashed}) estimates of the last three parameters are obtained based on the orbital inclination $90^\circ$ and $44^\circ$, respectively. The lower limit of the inclination angle is determined from the maximum measured projected separation of 0.042~pc ({\it dashed black} line). Combining the orbital inclination and the measured $f^{(4)}$, the companion mass and the pulsar orbital period and size are constrained to be about $(8,000-37,000)$~$M_\odot$, $(1.36-1.94)$~kyr, and $(0.012-0.025)$~pc, respectively.
\label{curves}}
\end{figure}

As the fourth derivative of the spin frequency ($f^{(4)}$) has a large uncertainty (see Table~\ref{solution}) compared to the other three lower derivatives we solve the orbital parameters as a function of $f^{(4)}$ by keeping $\dot \lambda$ at $\dot \lambda_{\rm mean}$, and obtain the other parameter uncertainties based on the measured uncertainty of $f^{(4)}$.
Our results are given in Table~\ref{solution} and shown in Figure~\ref{curves}. This indicates that in this model the pulsar is currently located near apastron of an extremely eccentric orbit ($e = 0.993(4)$)  that includes the cluster centre.  We note that while $e$, $\omega$, and $\lambda$ are independent of inclination angle, the derived companion mass, orbital size and projected separation all depend on the orbital inclination. However we can use the maximum possible measured projected separation of the pulsar from the cluster centre ($\approx 0.042$~pc) to impose a lower limit on the inclination angle of $i \approx 44^\circ$ (see Figure~\ref{curves}). The remaining possible inclination range $44^\circ < i < 90^\circ$ predicts that the mass of the companion, which would therefore be the mass at the cluster centre $M$, is between about $8,000-37,000$~$M_\odot$ (Figure~\ref{curves}). Using these parameters in Kepler's laws, we also constrain the orbital period ($P_{\rm orb} = 1.6(4)$~kyr), which is independent of the orbital inclination, indicating that the pulsar is in a wide orbit. 
Moreover, the orbital parameters show the location of the pulsar is very close to the sky plane (i.e. $\lambda \approx 180^\circ$ and $\omega \approx 0^\circ$). Thus, the radial distance of the pulsar from the cluster centre is approximately equal to the projected distance.
We further note that with these parameters the model-predicted $f^{(5)} = 0.7\times10^{-52}$~s$^{-6}$ obtained from Equation~\ref{f5} is consistent with the timing measured upper limit of $1(2)\times10^{-52}$~s$^{-6}$ (see Table~\ref{solution}). 

In principle the pulsar might be orbiting a cluster centre full of stars, however, combining this mass with the periastron separation of $(9-14)\times10^{-5}$~pc, where this range is due to the possible orbital inclinations, this leads to a stellar density of $3\times10^{15}$~$\rm M_\odot/$pc$^{3}$ compared with the low central density of the cluster of $\sim$$10^{5}$~$\rm M_\odot/$pc$^{3}$ derived from the early optical observations \citep{cgh+78}. Such an exceptional mass density can only be achieved if the pulsar is orbiting an intermediate-mass black hole (IMBH).

\subsubsection{Case II: Orbital results with a significant intrinsic spin-down of the pulsar}
\label{result_b}

As discussed in \citet{faa+11} the high $\gamma$-ray luminosity of the pulsar could indicate that a large fraction of the observed spin-down is intrinsic. Here we take account of this possibility in the orbit model. Although it is not possible to estimate the intrinsic spin-down, we can estimate a sensible range by using the $\gamma$-ray luminosity with the  assumption that the $\gamma$-ray emission is directly related to the intrinsic spin-down of the pulsar. Using the measured $\gamma$-ray flux density of $(1.60\pm0.17)\times10^{-11}$~erg$/$s$/$cm$^2$ given in \citet{aaa+15}, we estimate the $\gamma$-ray luminosity of the pulsar to be $L_\gamma = 1.1\times10^{35} f_\Omega (D/7.6)^2$~erg/s for a given $
\gamma$-ray beaming fraction of $f_\Omega$ and a pulsar distance of $D$ in kpc. The parameter $f_\Omega$ varies widely between different emission models for this pulsar \citep[$\sim$0.3--0.94, see][]{faa+11,jvh+14}, and for all $\gamma$-ray pulsars in general. Therefore, we assume the standard value of $f_\Omega \approx 1$ in this case. Then we can get the intrinsic spin-down ($\dot{f}_{\rm int}$) of the pulsar assuming a $\gamma$-ray emission efficiency of $\eta$ $(= L_\gamma/\dot{E}$, where $\dot{E} = -4\pi^2 I f \dot{f}_{\rm int}$ is the intrinsic spin-down energy of the pulsar) to be 

\begin{equation}
\label{fdot}
\dot{f}_{\rm int} = -1.5\times10^{-14} (f_\Omega/\eta) (D/7.6)^2~\rm s^{-2}
\end{equation}

\noindent
for an assumed pulsar moment of inertia of $I = 10^{45}$~g~cm$^2$. This leads to a dynamically induced spin-down of $\dot{f}_{\rm dyn} = \dot{f} - \dot{f}_{\rm int}$, which we can input to the orbital model to constrain the pulsar orbit and the companion mass as described before in Section~\ref{result_a}. Note that $\eta$ cannot be estimated without having a well-constrained $f_\Omega$ and a good estimate for the pulsar distance. Thus we express $\dot{f}_{\rm dyn}$ as a function of $\eta$ for the given pulsar distance of $7.6$~kpc \citep{kdi+03} and $f_\Omega=1$.

\begin{figure}
\begin{center}
\includegraphics[width=3.in]{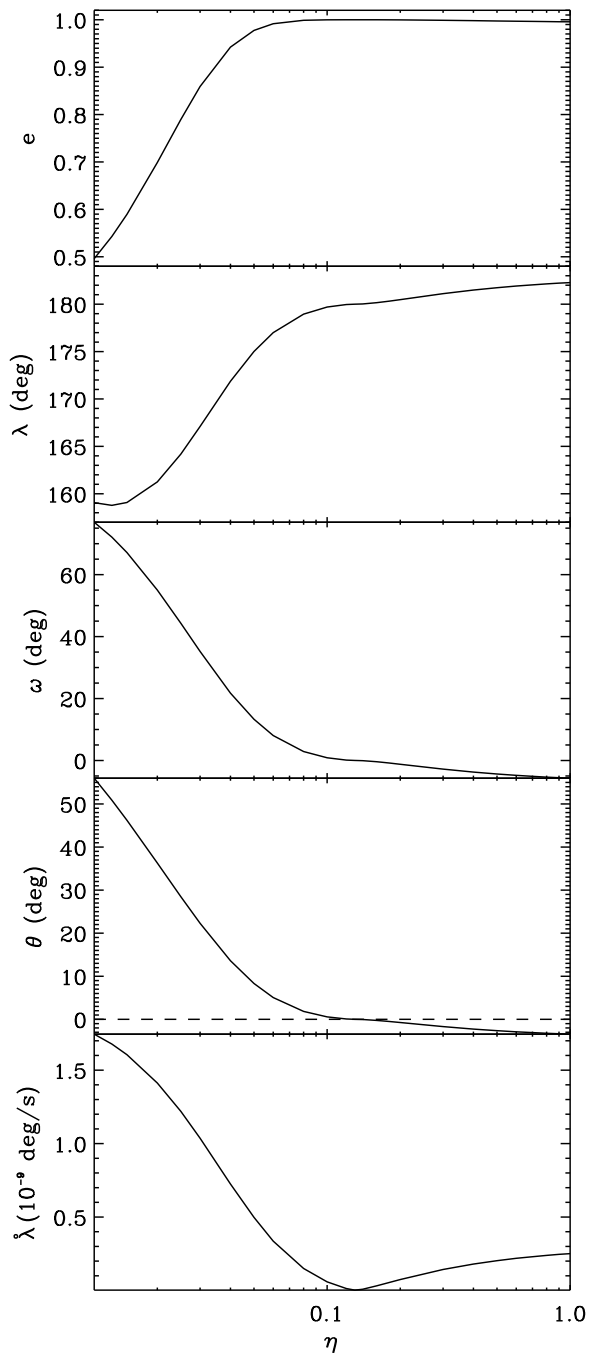}
\end{center}
\caption{
Constrained orbital parameters as a function of $\eta$ when taking account of the pulsar intrinsic spin-down $\dot{f}_{\rm int}$ (see Equation~\ref{fdot}) in the orbit model. Note that $\theta$ is the angle subtended by the pulsar at the centre with respect to the sky-plane through the cluster centre. The sign of $\theta$ and $\dot{f}_{\rm dyn}$ changes at the critical limit of $\eta \approx 0.13$ where $\dot f \approx \dot{f}_{\rm int}$.
\label{dyn_sol}}
\end{figure}

By following the same procedure given in Section~\ref{result_a} using the estimated $\dot{f}_{\rm dyn}$ instead of $\dot f$, we determine the orbital-inclination-independent parameters for the measured $f^{(4)}$ as a function of $\eta$ (see Figure~\ref{dyn_sol}). The calculated $\dot{f}_{\rm dyn}$ is negative, but it is possible for it to have positive values for small $\eta$ where $|\dot{f}| < |\dot{f}_{\rm int}|$. This sign variation corresponds to positive or negative acceleration of the pulsar towards our line-of-sight and can be understood in terms of the pulsar location in the orbit with respect to the sky-plane which passes through the cluster centre \citep[see Figure~1 in ][]{phi92}. The angle $\theta$ $(= \omega + \lambda - \pi)$, which is the angle subtended by the pulsar at the cluster centre with respect to the sky-plane through the centre, becomes negative (i.e. when the pulsar is located behind this particular sky-plane) when $\dot{f}_{\rm dyn}$ is negative (i.e. $\eta>0.13$ and the pulsar acceleration is positive), and vice-versa. This can be clearly seen in Figure~\ref{dyn_sol}.

\begin{figure*}
\begin{center}
\includegraphics[width=6in]{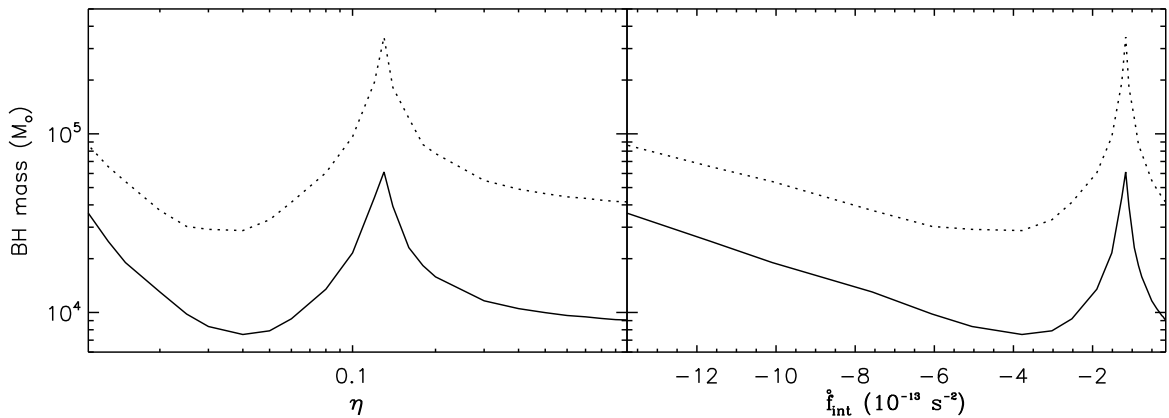}
\end{center}
\caption{
The derived central IMBH mass as a function of $\gamma$-ray efficiency ({\it left}) or pulsar intrinsic spin-down ({\it right}). The lower limit of the mass ({\it solid}) is obtained from the orbital inclination of $i=90^\circ$ and the possible upper limit ({\it dotted}) is obtained for $i=44^\circ$ using the fact that the possible maximum projected distance of the pulsar from the cluster centre should be $\sim$0.042~pc as described before in Figure~\ref{curves}.   
\label{dyn_mass}}
\end{figure*}

Using these parameters, we then derived the companion mass, or the central mass of the cluster, as a function of $\eta$. This result can also be represented as a function of $\dot{f}_{\rm int}$ using Equation~(\ref{fdot}) (see Figure~\ref{dyn_mass}). The maximum physically meaningful $\eta$ is unity, but we note that it could be greater than one due to an overestimated pulsar distance and $f_\Omega$. As described in Equation~\ref{fdot}, the lowest possible intrinsic spin-down of the pulsar is $\dot{f}_{\rm int} = -1.5\times10^{-14}$~s$^{-2}$ for $\eta=1$. 
The highest intrinsic spin-down occurs when the pulsar has the lowest $\eta$. According to \citet{aaa+13}, the lowest $\eta$ for $\gamma$-ray MSPs is estimated to be $0.013$, which is equivalent to $\dot{f}_{\rm int} \approx -1.2\times10^{-12}$~s$^{-2}$ (from Equation~\ref{fdot}). Therefore, we use a range of $\dot{f}_{\rm int}$ between $-1.3\times10^{-12}$~s$^{-2}$ and $-0.15\times10^{-13}$~s$^{-2}$ (i.e. $\eta$ is between $\sim$$0.011-1$ using Equation~\ref{fdot}) to estimate the central mass of the cluster. 
We also note that this range covers the estimated $\eta (\approx 0.03-0.2)$ for this pulsar based on different $\gamma$-ray emission models \citep[see][]{faa+11,jvh+14,aaa+13}.
Our results shown in Figure~\ref{dyn_sol}  and \ref{dyn_mass} indicate that the mass of the companion is M $>7,500$~M$_\odot$ and the orbit includes the cluster centre. Combining this mass with the periastron separation of the pulsar orbit, we calculate the central mass density of the cluster to be $>$$2\times10^9$~M$_\odot /$pc$^3$. As described in Section~\ref{result_a}, this extreme mass density is achieved only if the central object is an IMBH. We also note that in the scenario where $\dot{f}_{\rm int} = \dot f$, there is no $\dot{f}_{\rm dyn}$ and the acceleration along the line-of-sight is zero and therefore, the central source is unconstrained and the mass estimate is not  meaningful (see the resultant spike in Figure~\ref{dyn_mass} around $\eta = 0.13$). We use the same information and method as described in Section~\ref{result_a} to impose the upper limit of the IMBH mass and find that the minimum possible orbital inclination is still at $\sim 44^\circ$. Note that if we assume a large $\eta (>1)$, then $\dot{f}_{\rm int} \rightarrow 0$ and the model converts to Case I given in Section~\ref{result_a}, where there is a lower limit for the central IMBH of $\sim$8,000~M$_\odot$.

In summary, regardless of the contribution of the pulsar intrinsic spin-down $\dot{f}_{\rm int}$ to its observed value $\dot f$, the orbital model predicts that NGC 6624 contains an IMBH in the cluster centre and PSR B1820$-$30A is in a wide high-eccentricity orbit which passes around this central source.
The possibility of NGC 6624 hosting a central IMBH  has previously been discussed in several studies based on different globular cluster models \citep[see][]{bah76,cgh+78,pbk+14}.

\subsection{Pulsar in a Smaller orbit with a less massive companion}
\label{small}

As we found previously, the possibility of the pulsar existing in smaller orbits with low-eccentricities including other classes of companions such as 
main sequence stars, white dwarfs, neutron stars or stellar-mass BHs cannot be completely ruled out. 
\citet{pbk+14} used a globular cluster dynamical model to estimate the acceleration of the LMXB and pulsars, and found a lower limit of the mass of a central IMBH to be 19,000~M$_\odot$. 
They mentioned that a central concentration of dark remnants is more favourable than a central IMBH based on the stability of the triple behaviour of the LMXB. However, we argue that the LMXB experiences the gravity from both scenarios similarly, and thus cannot separate the two cases (see Section~\ref{dis} for details).
Therefore, we study the stability of the range of possible smaller orbits with low-mass-companions for PSR B1820$-$30A at its location in the cluster due to the presence of this 19,000~M$_\odot$ IMBH at the centre.

\begin{figure}
\begin{center}
\includegraphics[width=3.2in]{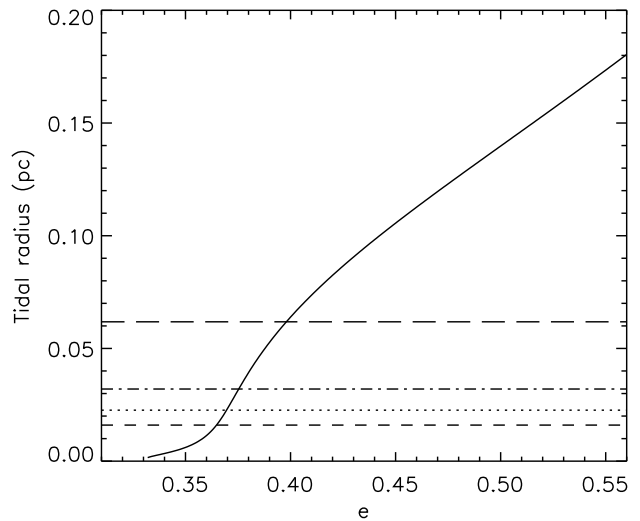}
\end{center}
\caption{
The tidal disruption radii for smaller orbit pulsar binary systems (i.e. $0.33\lesssim e\lesssim0.56$) due to the central IMBH derived from the globular cluster model given in \citet{pbk+14}. The separation of the pulsar from the cluster centre are marked by straight lines accordingly for $\theta=0^\circ$ ({\it dashed}), $45^\circ$ ({\it dotted}), $60^\circ$ ({\it dot-dashed}), and $75^\circ$ ({\it long-dashed}), where $\theta$ is the angle of the pulsar location measured from the sky plane through the cluster centre. 
All binary possibilities where the tidal radius is greater than the pulsar separation can be ruled out, remaining only valid systems within $0.33\lesssim e\lesssim0.40$. 
}
\label{tidal}
\end{figure}

We estimate the radii at which these binaries would tidally disrupt due to the central IMBH and determine if the pulsar separation from the cluster centre is smaller than this tidal radius. However, the true pulsar location in the cluster for this case is unknown. 
If we assume that the pulsar is located in the sky plane which passes through the cluster centre, then the separation between the pulsar and the cluster centre is $\sim0.016$~pc (i.e. the minimum measured projected distance between the pulsar and the cluster centre; see Section~\ref{model}). In this case we can rule out smaller orbit binary systems with $e\gtrsim 0.37$ as shown in Figure~\ref{tidal} for an edge-on orbit. 
We then vary the pulsar location in the cluster and investigate the stability of the orbit. We find that if the pulsar is located about $75^\circ$ off from the sky plane  (i.e. $\theta=75^\circ$), then the binary orbits with $e\gtrsim0.4$ would tidally disrupt (see Figure~\ref{tidal}). This indicates that we can rule out all the possible orbital systems with $e\gtrsim0.4$ for any of the given highly-probable pulsar locations in the cluster (i.e. $\theta<75^\circ$), leaving only very low-eccentricity binaries within a narrow valid region of  $0.33\lesssim e\lesssim0.4$.

However, as described in \citet{pbk+14}, it is likely that the other pulsars and the LMXB in this cluster are orbiting around the central IMBH. Therefore, it is more likely that PSR B1820$-$30A is also orbiting around the central IMBH and thus we consider that all the possible smaller orbits with the less massive companions are unlikely.

\begin{table*}
\begin{center}
\caption{
Summary of possible orbital models for PSR B1820$-$30A. Case I -- The pulsar orbits around the massive central source of the cluster and its intrinsic spin-down is negligible compared to the timing measured value. Case II -- This is similar to Case I, but for a significant intrinsic spin-down compared to the measurement. Case III -- The pulsar orbits with a less-massive companion in a smaller orbit. Based on the properties of the cluster with other pulsars and the LMXB, Case I and II are the most probable orbital models for the pulsar than Case III (see Section~\ref{model} for details).   
}
\label{summary}
\begin{tabular}{lcccccc}
\hline
\multicolumn{1}{l}{} &
\multicolumn{1}{c}{$e$} &
\multicolumn{1}{c}{$\lambda$} &
\multicolumn{1}{c}{$\omega$} &
\multicolumn{1}{c}{Orbit size} &
\multicolumn{1}{c}{Orbital period} &
\multicolumn{1}{c}{Companion mass} \\
\multicolumn{1}{l}{} &
\multicolumn{1}{c}{} &
\multicolumn{1}{c}{(deg)} &
\multicolumn{1}{c}{(deg)} &
\multicolumn{1}{c}{(pc)} &
\multicolumn{1}{c}{(kyr)} &
\multicolumn{1}{c}{(M$_\odot$)} \\
\hline
Case I & 0.993 & 183 & 353 & 0.01$-$0.025 & 1.3$-$2 & 8,000$-$37,000 \\
Case II & $>$0.5 & 159$-$183 & 355$-$420 & 0.02$-$0.04 & 0.5$-$3 & $>$7,500 \\
Case III & 0.33$-$0.4 & 160$-$175 & 73$-$78 & 0.0001$-$0.0004 & 0.7$-$1.2 & 0.5$-$2 \\
\hline
\end{tabular}
\end{center}
\end{table*}

\section{Gravitational model for NGC 6624}
\label{dyn}

In order to compare our results with the results given in \citet{pbk+14}, we use their globular cluster model with our timing derived central IMBH\footnote{Note that we re-derived their equations and followed our own procedure to obtain results in the analysis. We noticed that our results are slightly different from those given in \citet{pbk+14}, and the reason for this discrepancy is not understood. We contacted the authors in that study, but could not come to a conclusion about the different results in the two analyses.}.
This model uses a mass profile which is derived from the surface density profile for the given best fit parameters and a total cluster mass of $2.5\times10^5$~M$_\odot$, excluding the mass of the central IMBH. We place the pulsar timing and dynamically derived central IMBH (i.e. $\sim$$10,000$~M$_\odot$ IMBH from Section~\ref{result_a} ) in the cluster and use Equation~(6) in their study to estimate the acceleration of the pulsar at the current position in the orbit (derived from the orbital parameters in Table~\ref{solution}). 
The model-predicted pulsar acceleration is then estimated to be $\sim2\times10^{-8}$~yr$^{-1}$, which is consistent with the pulsar acceleration derived from our measured value of $\dot f$ given in Table~\ref{solution}.

\begin{figure}
\begin{center}
\includegraphics[width=3.3in]{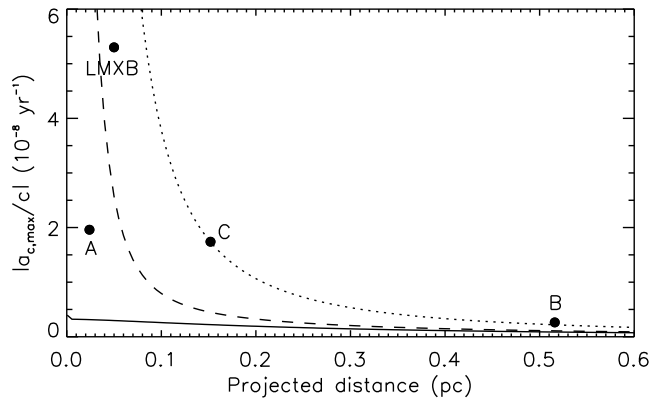}
\end{center}
\caption{
Cluster model-predicted maximum acceleration as a function of projected distance from the centre. Different curves represent the acceleration for different IMBH masses in the centre: no IMBH ({\it solid}), $10,000$~M$_\odot$ ({\it dashed}), and $60,000$~M$_\odot$ ({\it dotted}). The measured accelerations of PSRs B1820$-$30A, B1820$-$30B, and B1820$-$30C, and LMXB 4U 1820$-$30 are marked appropriately. To be consistent with the measured acceleration of these sources, the cluster model requires the minimum mass of the central IMBH to be $60,000$~M$_\odot$.
\label{peuten}}
\end{figure}

We then use this model to predict the acceleration of the other two pulsars and the LMXB in the cluster that have measured accelerations derived from their period derivatives \citep[see][]{pbk+14}. Since the current 3D position of these sources in the cluster are not available, like our measurement for PSR B1820-30A, only the maximum acceleration along the line-of-sight can be predicted (see Figure~\ref{peuten}). To get consistent maximum acceleration estimates with the measured source accelerations, we find that the minimum mass of the central IMBH is increased to $\sim$60,000~$M_\odot$. This large mass limit is consistent with the timing derived mass when the pulsar intrinsic spin down $\dot{f}_{\rm int}$ lies between about $-1\times10^{-12}$~s$^{-2}$ and $-7.6\times10^{-14}$~s$^{-2}$ (i.e. $0.08\leqslant\eta\leqslant0.2$) as shown in Figure~\ref{dyn_mass}. This $\eta$ range is consistent with that measured for the pulsar based on different $\gamma$-ray emission models \citep[see][]{jvh+14,aaa+13}.

We further note that our model depends on the density profile of the globular cluster \citep[see][for more details]{pbk+14}. We assumed in the model that the mass-to-light ratio of the cluster is unity and also the orbits of all pulsars and the LMXB are in the same plane through the cluster centre and edge-on to our line-of-sight. Therefore, we note that this model-estimated minimum mass ($\sim$60,000~M$_\odot$) of the central IMBH depends on various assumptions and parameter values used in the cluster model.

\section{Discussion}
\label{dis}

We have determined the orbital parameters and the companion mass of PSR B1820$-$30A using the measured higher-order rotational frequency derivatives through pulsar timing. 
Our results are consistent with the pulsar being in either a high-eccentricity larger orbit ($e\gtrsim0.9$) around the central IMBH or in a low-eccentricity smaller orbit ($0.33\lesssim e\lesssim0.4$) with a low-mass companion. 
However, the properties and the measurements of the cluster and the other sources near the centre are consistent with a central IMBH (see Figure~\ref{peuten}), and thus PSR B1820$-$30A is highly likely orbiting around a central IMBH, ruling out the smaller orbit possibilities. 
Regardless of the contribution of the pulsar intrinsic spin-down to the observed value shown in  timing measurements, we deduce that NGC 6624 contains an IMBH with a mass of M $> 7,500$~M$_\odot$ at the centre and that the pulsar orbits around it (see Table~\ref{summary}). This is the first evidence and mass constraint of a central BH in a globular cluster made using the timing measurements of pulsars directly combined with orbital dynamics. This mass estimate is consistent with that of central IMBHs in other Galactic globular clusters constrained through photometry of stars with kinematic models \citep{lkn+11,lkg+13,lgb+15}. 
The derived orbital parameters and the mass limits indicate that the inclination of the pulsar orbit must be between about $44^\circ<i<90^\circ$ and the orbital period is in the range $1.36-1.94$~kyr, where the error on the period comes from the uncertainty of our measured $f^{(4)}$.  The extremely high eccentricity of the orbit ($e = 0.993(4)$) was likely to be the result of a dynamical interaction deep in the cluster centre. This is somewhat similar to high eccentricities that are found in some globular cluster binary pulsars \citep[see][]{drk+15,frg07,lfrj12}. The current likelihood of a strong gravitational interaction is sufficiently low, both at periastron and apastron,  to suggest that the current orbit is stable (see Appendix~\ref{time}). 
We found that the model-predicted $f^{(5)}$ calculated using the best-fit orbital parameters of the pulsar is consistent with the timing measured upper limit. 
In the future with about four years of more LT observations, we will be able to obtain a measurement for $f^{(5)}$ with an accuracy of about 2-$\sigma$, including improvements in the other rotational derivative measurements, to constrain all four parameters in the orbital model independently, leading to a better IMBH mass estimate. Combining observations from other telescopes with the LT observations, we will be able to reach the required accuracy of $f^{(5)}$ in a shorter period than mentioned above. We also considered the possibility of the cluster containing large amount of dark remnants in the inner region instead of a central IMBH as argued in \citet{pbk+14}. 
As used in \citet{pbk+14} for this dark remnant scenario (see Section~4.4 therein), we considered the inner slope of the mass density profile to be 1.7 in the globular cluster model given here in Section~\ref{dyn}. We estimated the total mass interior to the measured projected distance of the pulsar ($\approx 0.02$~pc) to be about $4,000$~M$_\odot$, again ruling out these possible low eccentricity orbits.

Globular clusters have long been proposed to host central IMBHs, but hitherto the evidence has not been conclusive \citep{mh02,scm+12a}. 
\citet{ngb08} claimed a central IMBH in the Galactic globular cluster $\omega$ Centauri by fitting isotropic cluster dynamical models to observed radial velocity dispersions, but its mass estimate is controversial due to difficulties in obtaining the position of the cluster centre \citep[see][]{va10,ngk+10}. 
\citet{lkn+11} claimed a possibility of a central IMBH in the Galactic globular cluster NGC 6388, but its mass estimate is also controversial \citep{lmo+13} due to difficulties in obtaining accurate velocity dispersions of the cluster stars \citep{lgb+15} near the centre because of the confusion with the large number of cluster stars along the line-of-sight which are not in the centre.   
This claim also relied on a dynamical model with the cluster mass distribution. In contrast, our detection in NGC 6624 is derived directly from the measured dynamics of PSR B1820$-$30A and its location in the cluster and is therefore independent of cluster model parameters and geometry assumptions used in previous studies \citep{ngk+10,lkn+11,pbk+14}.

Based on the stability of the assumed triple system scenario of LMXB 4U 1820$-$30, \citet{pbk+14} found that a flyby dark remnant with a mass of $<6$~M$_\odot$ could explain its observed negative period derivative. However, they also mentioned that the chance of experiencing a similar flyby event by all other three pulsars is highly unlikely.
Then they argue that the massive IMBH derived from their globular cluster model implies that the observed period derivative of this source is a unique event with a short duration and the triple behaviour would be destroyed by tidal forces very quickly. In addition, they argue that if the observed period derivatives of all three pulsars and the LMXB are dynamically induced by the strong gravitational force exerted within NGC 6624, then it is necessary to have an extended concentration of non-luminous mass in the central region; e.g. dark subsystem of BHs, neutron stars, and white dwarfs. They suggested that an order of 70,000~${\rm M_\odot}$ dark remnants could be concentrated within the break radius of 0.22~pc in the cluster, resulting in a mass density of $10^6$~${\rm M_\odot}$/pc$^3$. 
They mentioned that the dark remnant scenario is more favourable than the central IMBH. We modified the globular cluster model given in Section~\ref{dyn} according to the parameters that \citet{pbk+14} used in their study of the dark remnant scenario and found that the total dark mass interior to the radius of the LMXB is about 14,000~${\rm M_\odot}$.
The triple system experiences the gravity of this large central dark mass as a single massive object located at the cluster centre, which is similar to a central IMBH. 
We therefore argue that there is no reason to favour the dark remnant scenario over a central IMBH.
The idea that the centre of this cluster contains dark remnants has also been discussed in \citet{gcl+92}. They suggested that a dark mass of $56,000$~${\rm M_\odot}$ is bounded within a central radius of $0.3$~pc, resulting in a mass density of $5\times10^5$~${\rm M_\odot}$/pc$^3$.
In contrast, our timing-derived orbital results find that a mass of $\sim10,000$~${\rm M_\odot}$ is bounded within even a smaller central radius of $\sim10^{-4}$~pc (see Section~\ref{result_a}), providing a mass density of $\sim10^{15}$~${\rm M_\odot}$/pc$^3$. Such an extremely high central mass density can only be achieved by an IMBH and therefore, the dark remnant scenario is unlikely. \citet{ngb08} found that a mass of $\sim10,000$~${\rm M_\odot}$ is bounded to a central radius of 0.05~pc of the globular cluster $\omega$ Centauri and claimed that it is more favourable to have an IMBH rather than a concentration of dark remnants.

The spherically symmetric mass accretion onto a central BH can produce X-ray and radio continuum emission \citep{bon52}. 
The mass limits on proposed IMBHs have been previously investigated through comparing the expectations from Bondi-accretion and X-ray and radio observations \citep[e.g.][]{ms08,mac05,scm+12a}. 
By following \citet{scm+12a} with the assumption of a 3\% Bondi-accretion rate at $10^4$~K and a gas number density of $0.2$~cm$^{-3}$, we estimate the expected radio luminosity to be between $8.3\times10^{30}$~erg/s (for an adiabatic process) and $2\times10^{32}$~erg/s (for an isothermal process) for an IMBH of mass $60,000$~${\rm M}_\odot$ (i.e the minimum possible IMBH mass, see Section~\ref{dyn}). The details of the calculations are given in Appendix~\ref{bondiacc}. 
\citet{mfr+04} reported the radio emission around the central region of NGC 6624, in particular around the LMXB. Their observations indicate that the flux density at the cluster centre is approximately four times the rms of their image \citep[i.e. $4\times0.015$~mJy -- see Figure~3 of][]{mfr+04}. Assuming this flux density, we calculate the measured radio luminosity at the centre is about $3.5\times10^{28}$~erg/s for a flat emission spectrum, which is about two orders of magnitude less than the luminosity expected  from the Bondi-accretion for an adiabatic case (i.e. $\gamma=5/3$). As mentioned in \citet{mfr+04}, the measured high frequency radio emission is attributed with the LMXB and not with PSR 1820$-$30A. We also note that the spatial separation between the radio emission contours and the cluster centre indicates that this emission is unlikely to be associated with the central IMBH.
If Bondi-accretion is active in the centre, then the model-predicted and the measured luminosities suggest that the percentage assigned to mass accretion or the gas density, or both, must be less than the assumed values in the calculation. We also note that the other assumptions may have affected the mass limit and the luminosity estimation (see Appendix~\ref{bondiacc}).

The detection of IMBHs is important for understanding the missing link between stellar mass black holes \citep{scm+12,psm14} and super massive black holes \citep{vol10}. It is generally thought that they could be formed by the direct collapse of very massive primordial stars \citep{mr01b}, or successive mergers of stellar-mass BHs \citep{mh02} and runaway collisions in dense young star clusters \citep{pbh+04}. 
Our dynamical mass measurement provides the clearest dynamical evidence yet for an IMBH with a reasonably well constrained mass and its location in a globular cluster provides important input to our understanding of how IMBHs and the clusters themselves, form and evolve \citep{pbh+04,mh02}. It also highlights the value of finding multiple pulsars in globular clusters, in particular close to their centres, where other techniques have difficulties with resolution, and the importance of undertaking long-term timing studies to understand their dynamics and potentially identifying and measuring the masses of further IMBHs.

\noindent

\section*{Acknowledgments}
Pulsar research at the Jodrell Bank Centre for Astrophysics and the observations using the Lovell Telescope are supported by a consolidated grant from the STFC in the UK. The Nan{\c c}ay radio Observatory is operated by the Paris Observatory, associated to the French Centre National de la Recherche Scientifique (CNRS) and to the Universit{\'e} d'Orl{\'e}ans. CGB acknowledges support from the European Research Council under the European Union's Seventh Framework Programme (FP/2007- 2013) / ERC Grant Agreement nr. 337062 (DRAGNET; PI Jason Hessels). MK acknowledges financial support by the European Research Council  for the ERC Synergy Grant BlackHoleCam under contract  no. 610058. We thank James Miller-Jones for useful discussions about the calculation of the expected X-ray and radio luminosities due to Bondi-accretion onto a BH.

\bibliography{psrrefs,modrefs,journals,0737Ack}
\bibliographystyle{mnras}

\appendix
\section{Orbital dynamical model}
\label{app}

The analytical expressions derived in \citet{jr97} for the `four-derivative' solution are given as a function of orbital parameters;  eccentricity $e$, the longitude of periastron $\omega$ (measured from the ascending node), the longitude of the reference epoch $\lambda$ (measured from the periastron), and $\dot \lambda$. We present these derivatives below, where the derivation of $f^{(5)}$is new to this work. We note that $\ddot \lambda = 2(A'/A)\dot{\lambda}^2$ is a useful term in the derivation and this can be shown by using the orbital dynamics given in the Appendix of \citet{bjs+16}.

\begin{equation}
\label{f2}
\ddot f = \frac{B \dot\lambda \dot f}{A^2\sin(\lambda + \omega)},
\end{equation}    

\begin{equation}
\label{f3}
f^{(3)} = \frac{C \dot\lambda^2 \dot f}{A^2\sin(\lambda + \omega)},
\end{equation}

\begin{equation}
\label{f4}
f^{(4)} = \frac{D \dot\lambda^3 \dot f}{A^2\sin(\lambda + \omega)},
\end{equation}

\begin{equation}
\label{f5}
f^{(5)} = \frac{E \dot\lambda^4 \dot f}{A^2\sin(\lambda + \omega)},
\end{equation}

\noindent
where,

\begin{equation}
A = 1 + e\cos\lambda,
\end{equation}

\begin{equation}
B = 2AA'\sin(\lambda+\omega) + A^2\cos(\lambda+\omega),
\end{equation}

\begin{equation}
C = B' + \frac{2BA'}{A},
\end{equation}

\begin{equation}
D = C' + \frac{4CA'}{A},
\end{equation}

\begin{equation}
E = D' + \frac{6DA'}{A},
\end{equation}

\noindent
with a prime indicating a derivative with respect to $\lambda$.

In order to reduce the number of variables, we slightly modified our main equations. We find $\dot \lambda$ from Equation~\ref{f2} and substitute in Equation~\ref{f3} and \ref{f4} to get new expressions

\begin{equation}
\label{ldot}
\dot{\lambda} = \frac{A^2 \sin(\lambda+\omega) \ddot{f}}{B \dot{f}}
\end{equation}

\begin{equation}  
\label{f3_new}
f^{(3)} = \frac{CA^2 \sin(\lambda+\omega)(\ddot f)^2}{B^2 \dot f}
\end{equation}

\begin{equation}  
\label{f4_new}
f^{(4)} = \frac{D[A^2 \sin(\lambda+\omega)]^2(\ddot f)^3}{B^3 (\dot{f})^2}.
\end{equation}

\noindent
We note that the unknown parameters in these two higher derivative equations are $\lambda$, $\omega$, and $e$. 

The orbital parameters ($e$, $\omega$, $\lambda$, and $\dot \lambda$) are then used to determine the orbital size and the mass of the companion as a function of the orbital inclination. These expression are given in \citet{jr97} as

\begin{equation} 
\label{asin}
a \sin (i) = -\frac{\dot f c A^2}{f(1-e^2)\sin(\lambda+\omega) \dot \lambda^2}~{\rm and}
\end{equation}

\begin{equation} 
\label{kk}
\frac{G M^3}{(m+M)^2} = -\left( \frac{\dot f c}{f\sin(i)\sin(\lambda+\omega)}\right)^3 \left( \frac{A^2}{\dot \lambda^4}\right),
\end{equation}

\noindent
where $a$ is the semi-major axis of the orbit, $i$ is the orbital inclination, $m$ is the pulsar mass (i.e. $m \approx 1.4$~M$_\odot$), $M$ is the mass of the companion, $G$ is the Gravitational constant, and $c$ is the speed of light. 
If the companion mass is large compared to the pulsar mass (i.e. $m << M$), we can simplify the above equation as

\begin{equation}
\label{msin}
M(\sin i)^3 = -\left( \frac{\dot f c}{f\sin(\lambda+\omega)}\right)^3 \left( \frac{A^2}{G\dot \lambda^4}\right).
\end{equation}

Assuming the pulsar orbits around the cluster centre, its projected distance from the centre can be found using the constrained orbital parameters as,

\begin{equation}
\label{r_perp}
R_\perp = \sqrt{Y^2 + Z^2}
\end{equation}

\noindent
where,

\begin{equation}
\label{yeqn}
Y = r(\cos\lambda\cos\omega - \sin\lambda\sin\omega)
\end{equation}

\begin{equation}
\label{zeqn}
Z = -r\cos i(\cos\lambda\sin\omega + \sin\lambda\cos\omega).
\end{equation}

\noindent
The radial distance of the pulsar in the orbit is 

\begin{equation}
\label{r_dis}
r = \frac{a(1-e^2)}{1+e\cos\lambda}.
\end{equation}

\section{Strong encounter timescale}
\label{timescale}

We consider the strong encounters of the pulsar with other stars during its orbital motion in the globular cluster environment. By definition, the timescale for strong encounters can be expressed as

\begin{equation} 
\label{time}
t_s \approx 4\times10^{12} \left( \frac{V}{10~{\rm km~s^{-1}}} \right)^3 \left( \frac{m}{{\rm M_\odot}} \right)^{-2} \left( \frac{n}{{\rm 1~pc^{-3}}} \right)^{-1}~{\rm yr}
\end{equation}

\noindent
where $V$ is the velocity of the pulsar,  $m$ is the average stellar mass in the cluster, and $n$ is the stellar density \citep[see Chapter~3.2.1 in][]{sg06}. 
Using the pulsar periastron and apastron velocities of 940~km~s$^{-1}$ and 3~km~s$^{-1}$, respectively, derived from the orbital parameters given in Case I combined with an average mass of 1~M$_\odot$ for all stars and a typical globular cluster stellar density of $10^6$~pc$^{-3}$, the strong encounter timescales are calculated to be $3.3\times10^{9}$~kyr and $130$~kyr at periastron and apastron, respectively.

We also considered Case II and found that for any given $\dot{f}_{\rm int}$ (except for the critical case of $\dot f \approx \dot{f}_{\rm int}$), the orbital period is smaller than $t_s$ at both periastron and apastron.

\section{Spherically symmetric accretion onto an IMBH}
\label{bondiacc}

As described in \citet{bon52}, a spherically symmetric accretion onto a BH can be written as

\begin{equation}
\dot{M}_{\rm B} = \pi G^2 M_{\rm BH}^2 \left( \frac{\rho}{c_{\rm s}^3} \right) \left( \frac{2}{5-3\gamma} \right)^{(5-3\gamma)/2(\gamma-1)}
\end{equation} 

\noindent
where $\gamma$ varies from 1 (for isothermal case) to $5/3$ (for adiabatic case), $\rho$ is the gas density, $c_{\rm s} = (\gamma kT/\mu m_{\rm p})^{1/2}$ is the sound speed of the gas, with the proton mass $m_{\rm p}$, and the mean mass per particle of gas $\mu$ in units of $m_{\rm p}$. As given in \citet{pel05} and \citet{scm+12a}, we assume $\mu=0.62$, $\rho=0.2m_{\rm p}$~kg/cm$^{3}$, and a 3\% of the Bondi-rate contribution to the mass accretion $\dot M$ for the gas at $10^4$~K (i.e. $\dot{M} = 0.03\dot{M}_{\rm B}$). Then the X-ray luminosity can be calculated as $L_{\rm X} = \epsilon \dot{M} c^2$, where $\epsilon = 0.1((\dot M/ \dot{M}_{\rm edd})/0.02)$ is the radiative efficiency. Here $\dot{M}_{\rm edd}$ is the mass loss at Eddington limit and we assume a 10\% efficiency on the Eddington luminosity, i.e. $\dot{M}_{\rm edd} = L_{\rm edd}/0.1 c^2$ \citep[see][]{pel05}.

\begin{figure}
\begin{center}
\includegraphics[width=3.in]{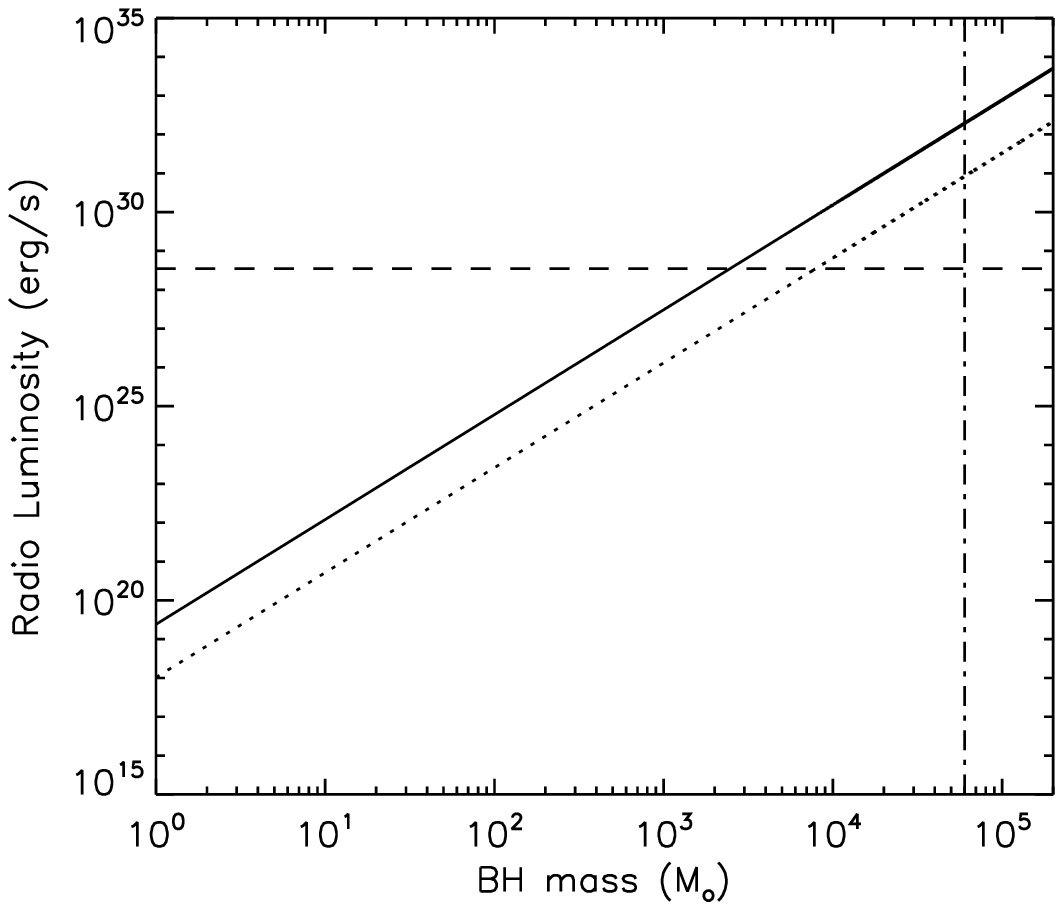}
\end{center}
\caption{
Expected radio luminosity due to mass accretion onto the central IMBH as a function of BH mass. The {\it solid} and {\it dotted} lines give the solutions for isothermal ($\gamma=1$) and adiabatic ($\gamma=5/3$) cases, respectively. The {\it dashed} line shows the calculated luminosity at the cluster centre based on the radio flux density reported in \citet{mfr+04}. The {\it dot-dashed} line represents the $60,000$~M$_\odot$ IMBH. To be consistent with the radio flux density, the required central IMBH mass is $\sim$8,000~M$_\odot$ (for $\gamma=5/3$).
\label{flux}}
\end{figure}

We then use the form of the X-ray--radio BH fundamental plane given in \citet{scm+12a}, 
\begin{equation}
\log L_{\rm X} = 1.44\log L_{\rm R} - 0.89\log M_{\rm BH} - 5.95,
\end{equation}

\noindent
to estimate the expected radio luminosity $L_{\rm R}$ for a given IMBH mass. 
By substituting all the typical assumptions and constants given above, we can obtain a useful expression for the radio luminosity as a function of the BH mass $M_{BH}$, in units of solar mass, as

\begin{equation} 
L_{\rm R} \approx \left[ 10^{31} \frac{(\eta n)^2}{\gamma^3} M_{\rm BH}^{3.89}  \left( \frac{2}{5-3\gamma} \right)^{(5-3\gamma)/(\gamma-1)} \right]^{0.695}~{\rm erg/s},
\end{equation}

\noindent
where $\eta$ and $n$ are the percentage of the Bondi-rate contribution to the mass accretion and the gas density in cm$^{-3}$, respectively.
By using this expression, we estimate the expected radio luminosity as a function of $M_{\rm BH}$ for the two cases $\gamma=5/3$ and $\gamma=1$ as shown in Figure~\ref{flux}.

\end{document}